\documentclass[
    ,final            
  ]
  {aipproc}

\layoutstyle{6x9}

\begin{document}

\title{Partonic substructure of nucleons and nuclei
with dimuon production}



\author{J.C.~Peng}{
  address={University of Illinois at Urbana-Champaign, Urbana, IL 61801}
}


\begin{abstract}
Dimuon production has been studied in a series of fixed-target
experiments at Fermilab during the last two decades. Highlights
from these experiments, together with recent
results from the Fermilab E866 experiment, are presented. Future
prospects for studying the parton distributions in the nucleons
and nuclei using dimuon production are also discussed.
\end{abstract}

\maketitle


\section{Introduction}

During the last two decades, a series of fixed-target dimuon
production experiments (E772, E789, E866) have been carried out
using 800 GeV/c proton beam at Fermilab.
At 800 GeV/c, the dimuon data contain Drell-Yan continuum
as well as quarkonium productions
(J/$\Psi$, $\Psi^\prime$, and $\Upsilon$ resonances). The Drell-Yan
process and quarkonium productions often provide complementary
information, since Drell-Yan is an electromagnetic process
involving quark-antiquark annihilation while the quarkonium production
is a strong interaction process dominated by gluon-gluon fusion
at this beam energy.

The Fermilab dimuon experiments covers a broad range of physics
topics. The Drell-Yan data have provided information on the
antiquark distributions in the nucleons~\cite{pat92,hawker98} 
and nuclei~\cite{alde90,vasiliev99}.
These results showed the surprising results that the
antiquark distributions in the
nuclei are not enhanced~\cite{alde90,vasiliev99},
contrary to the expectation of models which predict nuclear
enhancement of meson clouds. Later, the measurement of the 
Drell-Yan cross section ratios $p+d/p+p$ clearly established the
flavor asymmetry of the $\bar d$ and $\bar u$ distributions in
the proton, and the $x$-dependence of this
asymmetry was determined~\cite{hawker98}.
Pronounced nuclear dependences of quarkonium productions were 
observed for J/$\Psi$, $\Psi^\prime$, and $\Upsilon$ resonances
~\cite{alde91a,leitch95}.
The nuclear dependence of Drell-Yan cross
sections has also provided information on
the energy loss of quarks traversing the nucleus~\cite{vasiliev99,garvey03}.
In addition, the decay angular distributions for Drell-Yan~\cite{pat99,
zhu07}, J/$\Psi$~\cite{chang03}, and $\Upsilon$ resonances~\cite{brown01}
have been measured.

In this article we first discuss the subject of
the flavor structure of the sea quark distributions in the nucleons.
The observation of a striking flavor asymmetry of the nucleon sea, inspired
by the seminal work of Tony Thomas~\cite{thomas83}, has
profound implications on our current knowledge on the parton
substructures in the nucleons. Some recent
results from the dimuon production experiments are then presented.
Finally, prospects for future experiments to study flavor structures of
the nucleons and nuclei will be discussed.

\section{FLAVOR STRUCTURE OF LIGHT-QUARK SEA}

The earliest parton models assumed that the proton sea was flavor symmetric,
even though the valence quark distributions are clearly flavor asymmetric.
The similar
masses for the up and down quarks suggest that the nucleon sea should be nearly
up-down symmetric.
The issue of the equality of $\bar u$ and $\bar d$ was first
encountered in the Gottfried integral~\cite{gott},
given as
\begin{equation}
I_G = \int_0^1 \left[F^p_2 (x) - F^n_2 (x)\right]/x~ dx =
{1\over 3}+{2\over 3}\int_0^1 \left[\bar u_p(x)-\bar d_p(x)\right]dx,
\label{eq1}
\end{equation}
where $F^p_2$ and $F^n_2$ are the proton and neutron structure functions.
Under the assumption of a $\bar u$, $\bar d$ flavor-symmetric sea in
the nucleon, the Gottfried Sum Rule~\cite{gott}, $I_G
= 1/3$, is obtained.
The most accurate measurement~\cite{nmc91} of the Gottfried integral gives 
$ 0.235\pm 0.026$, significantly below 1/3. This surprising result has
generated much interest. It is remarkable that, already in 1983,
Tony Thomas predicted a large excess of $\bar d$ to $\bar u$ as a
direct evidence for a pionic component in the nucleon. In his 
paper~\cite{thomas83}, Tony commented on ``the lack of 
experimental information
about the shapes of $s(x)$, $\bar s(x)$ and $\bar d(x) - \bar u(x)$'',
and concluded his paper ``with a plea for better measurements of these 
three quantities in the free proton''.

The shape of the $\bar d(x) - \bar u(x)$ was later measured in 
proton-induced Drell-Yan and semi-inclusive DIS experiments.  
At Fermilab, the
E866/NuSea~\cite{hawker98} Collaboration
measured the DY cross section ratios for $p + d$ to that of $p + p$:
\begin{equation}
\sigma_{DY}(p+d)/2\sigma_{DY}(p+p) \simeq
\left[1+\bar d(x)/\bar u(x)\right]/2.
\label{eq:2}
\end{equation}

\noindent This ratio was found to be significantly
different from
unity for $0.015 <x< 0.35$, showing an excess of $\bar d$ 
to $\bar u$ over an appreciable range in $x$.

Many theoretical models, including meson-cloud model, chiral-quark
model, Pauli-blocking model, instanton model, chiral-quark soliton
model, and statistical model, have been proposed to explain the $\bar
d/ \bar u$ asymmetry. Details of these various models can be found in
some review articles~\cite{kumano98,garvey02}.
These models also have specific predictions for
the spin structure of the nucleon sea~\cite{peng03}. In the 
meson-cloud model, for example,
a quark would undergo a spin flip upon an
emission of a pseudoscalar meson ($
u^\uparrow \to \pi^\circ (u \bar u, d \bar d) + u^\downarrow,~u^\uparrow
\to \pi^+ (u \bar d) + d^\downarrow,~u^\uparrow \to K^+ + s^\downarrow$,
etc.). The antiquarks ($\bar u, \bar d, \bar s$) are
unpolarized ($\Delta \bar u = \Delta \bar d = \Delta \bar s = 0$)
since they reside in spin-0
mesons. The strange quarks ($s$), on the other hand, would have a negative
polarization since the up valence quarks in the proton
are positively polarized and
the $u^\uparrow \to K^+ + s^\downarrow$ process would lead to an excess
of $s^\downarrow$. By considering a vector meson ($\rho$) cloud,
non-zero $\bar u, \bar d$ sea quark polarizations with
$\Delta \bar d - \Delta \bar u > 0$ were
predicted~\cite{fries98,cao01}.

The Pauli-blocking model~\cite{steffens} predicts that an excess of
$q^\uparrow (q^\downarrow)$ valence quarks
would inhibit the creation of a pair of $q^\uparrow \bar q^\downarrow$
($q^\downarrow \bar q^\uparrow$) sea quarks. Since the polarization
of the $u$($d$) valence quarks is positive (negative), this
model predicts a positive (negative) polarization for the $\bar u
(\bar d)$ sea $(\Delta \bar u > 0 > \Delta \bar d)$.

In the instanton model~\cite{dorokhov}, the
quark sea can result from a scattering of a
valence quark off a nonperturbative vacuum fluctuation of the gluon field,
instanton. The correlation between the sea quark helicity and the valence quark
helicity in the 't Hooft effective lagrangian (i.e. $u^\uparrow$ leads to a
$\bar d^\downarrow$) naturally predicts a positively (negatively) polarized
$\bar u (\bar d)$ sea. In particular, this model predicts~\cite{dorokhov01}
a large $\Delta \bar u, \Delta \bar d$ flavor asymmetry with
$\Delta \bar u > \Delta \bar d$, namely,
$\int_0^1 [\Delta \bar u(x) - \Delta \bar d(x)] dx = \frac{5}{3} \int_0^1
[\bar d(x) - \bar u(x)] dx.$

In the chiral-quark soliton model~\cite{diakonov96,wakamatsu98},
the polarized isovector distributions
$\Delta \bar u(x) - \Delta \bar d(x)$ appears in leading-order ($N_c^2$)
in a $1/N_c$ expansion, while the unpolarized isovector distributions
$\bar u(x) - \bar d(x)$ appear in next-to-leading order ($N_c$).
Therefore, this model predicts a large flavor asymmetry for the polarized sea
$[\Delta \bar u (x) - \Delta \bar d(x)] > [\bar d(x) - \bar u(x)]$.

In the statistical model~\cite{bourrely01}, the momentum 
distributions for quarks and
antiquarks follow a Fermi-Dirac distributions function characterized
by a common temperature and a chemical potential $\mu$ which depends on
the flavor and helicity of the quarks. It can be shown that
$\mu_{\bar q\uparrow} = - \mu_{q\downarrow}$ and 
$\mu_{\bar q\downarrow} = - \mu_{q\uparrow}$.
Together with the constraints of the valence quark
sum rules and inputs from polarized DIS experiments, this model leads
to the prediction that $\bar d > \bar u$ and
$\Delta \bar u > 0 > \Delta \bar d$.

Measurements of $\Delta \bar u(x)$ and $\Delta \bar d(x)$ are clearly of
great current interest. Both the HERMES~\cite{hermes} and the 
COMPASS~\cite{compass} collaborations
have reported results on the extraction of $\Delta \bar u(x)$ and
$\Delta \bar d(x)$ from polarized semi-inclusive DIS data. These
results show that $\Delta \bar u$, $\Delta \bar d$ are small, but with
large uncertainties. A recent global analysis~\cite{DSSV} of polarized
DIS and polarized $p-p$ interaction indicates that $\Delta \bar u(x) > 0$,
$\Delta \bar d(x) < 0$, and $|\Delta \bar u(x)| < |\Delta \bar d(x)|$.
This interesting result suggests that the sea-quark polarization is
flavor-asymmetric and of opposite sign compared to the unpolarized
case. Additional data are anticipated for $W$-boson production 
at RHIC~\cite{bunce00}.
The longitudinal single-spin asymmetry for $W$ production in polarized
$p+p$ collision is sensitive to $\Delta \bar u(x)$ and $\Delta \bar d(x)$.

While various theoretical models can describe the general trend of
the $\bar d / \bar u$
asymmetry, they all have difficulties~\cite{pauchy91,wally99} 
explaining the Fermilab E866
data at large $x$ ($x>0.2$), where $\bar d / \bar u$ drops below 1.  
However, the E866 large-$x$ data suffer from large statistical
uncertainties, and more precise measurements are needed. 
The 120 GeV Main Injector at Fermilab and the new 30-50 GeV 
proton accelerator, J-PARC, present opportunities for
extending the $\bar d/ \bar u$ measurement to larger $x$ ($0.25 < x < 0.7$).
For given values of $x_1$ and $x_2$ the DY cross section
is proportional to $1/s$, hence the DY cross section
at these lower energies are significantly larger than at 800 GeV.
A definitive measurement of the $\bar d/ \bar u$ over the region 
$0.25 < x < 0.7$ could be obtained for an upcoming experiment
E906~\cite{e906} at Fermilab and a proposed measurement~\cite{p04}
at J-PARC.

To disentangle the $\bar d / \bar u$ asymmetry from the possible
charge-symmetry violation effect~\cite{ma}, one could 
consider $W$ boson production in $p + p$ collision at RHIC.
The ratio of the
$p + p \to W^+ + X$ and $p + p \to W^- + X$ cross sections
is sensitive to $\bar d / \bar u$. An important feature of
the $W$ production asymmetry in $p + p$ collision
is that it is completely free from the assumption 
of charge symmetry~\cite{londergan}.
Another advantage is that it is
free from any nuclear effects. Moreover, the $W$ production is
sensitive to $\bar d/ \bar u$ flavor
asymmetry at a $Q^2$ scale of $\sim$ 6500 GeV$^2/$c$^2$, significantly
larger than all existing measurements. This offers the opportunity to
examine the QCD evolution of the sea-quark flavor asymmetry.
A recent study showed that $W$ asymmetry
measurements at RHIC could provide an independent determination
of $\bar d / \bar u$~\cite{yang09}.

While it is generally assumed that the gluon
distributions in the proton and neutron are identical, this assumption
has not been tested
experimentally. A possible mechanism for generating different gluon
distributions in the proton and neutron, as pointed out by Piller
and Thomas~\cite{piller}, is the violation of charge symmetry in the 
parton distributions in the nucleons~\cite{londergan}.
Unlike the electromagnetic Drell-Yan process, quarkonium
production is a strong interaction dominated by the subprocess of
gluon-gluon fusion at 800 GeV beam energy. Therefore, the
$\Upsilon$ production ratio, $\sigma(p+d \to \Upsilon)
/\sigma(p+p \to \Upsilon)$, is expected to
probe the gluon content in the neutron relative to that in the
proton~\cite{zhu08}. 
The $\sigma(p+d)/2\sigma(p+p)$ ratios for $\Upsilon$ production with 
800 GeV proton beam have been
reported recently~\cite{zhu08}, and they are consistent with unity, 
in striking contrast to the corresponding values for the Drell-Yan process.
The $\Upsilon$ data indicate that the gluon distributions in the
proton and neutron are very similar.
These results are consistent with no
charge symmetry breaking effect in the gluon distributions.

\section{Transverse Spin and Drell-Yan Process}

The study of the transverse momentum dependent (TMD) parton distributions
of the nucleon has received much attention in recent years
as it provides new perspectives on the hadron structure and
QCD~\cite{barone02}. These novel TMDs can be extracted from semi-inclusive
deep-inelastic scattering (SIDIS) experiments. Recent measurements of
the SIDIS by the HERMES~\cite{hermes05}
and COMPASS~\cite{compass05} collaborations
have shown clear evidence for the existence of the T-odd Sivers functions.
These data also allow the first determination~\cite{vogelsang05} of
the magnitude and flavor structure of the Sivers functions and the nucleon
transversity distributions.

The TMD and transversity parton distributions can also be probed in
Drell-Yan experiments. As pointed out~\cite{ralston79}
long time ago, the double
transverse spin asymmetry in polarized Drell-Yan, $A_{TT}$, is proportional
to the product of transversity distributions, $h_1(x_q)h_1(x_{\bar q})$.
The single transverse spin asymmetry, $A_N$, is sensitive to the
Sivers function~\cite{sivers90}, $f^\perp_{1T}(x)$ of
the polarized proton (beam or target).
Even unpolarized Drell-Yan experiments can be used to probe the TMD
distribution function, since the cos$2\phi$ azimuthal angular dependence
is proportional to the product of two Boer-Mulders functions~\cite{boer98},
$h^\perp_1(x_1) \bar h^\perp_1(x_2)$. A unique feature of the Drell-Yan
process is that, unlike the SIDIS, no fragmentation functions are involved.
Therefore, the Drell-Yan process provides an entirely independent technique
for measuring the TMD functions. Furthermore, the proton-induced Drell-Yan
process is sensitive to the sea-quark TMDs and can lead to flavor separation
of TMDs when combined with the SIDIS data. Finally, the intriguing
prediction~\cite{collins02} that the T-odd TMDs extracted from
DIS will have a sign-change for the
Drell-Yan process remains to be tested experimentally.

No polarized Drell-Yan experiments have yet been performed.
However, some information on the
Boer-Mulders functions have been extracted recently
from the azimuthal angular distributions in the unpolarized Drell-Yan
process. The general expression for the Drell-Yan
angular distribution is~\cite{lam78}
\begin{equation}
\frac {d\sigma} {d\Omega} \propto 1+\lambda \cos^2\theta +\mu \sin2\theta
\cos \phi + \frac {\nu}{2} \sin^2\theta \cos 2\phi,
\label{eq:eq1}
\end{equation}
\noindent where $\theta$ and $\phi$ are the polar and azimuthal decay angle
of the $l^+$ in the dilepton rest frame. Boer showed that the $\cos 2\phi$
term is proportional to the convolution of the quark and antiquark
Boer-Mulders functions in the projectile and target~\cite{boer99}.
This can be understood by noting that the Drell-Yan cross
section depends on the transverse spins of the annihilating quark and
antiquark. Therefore, a correlation between the transverse spin and
the transverse momentum of the quark, as represented by the Boer-Mulders
function, would lead to a
preferred transverse momentum direction.

Pronounced $\cos 2 \phi$ dependences
were indeed observed in the NA10~\cite{falciano86} and E615~\cite{conway89}
pion-induced Drell-Yan experiments, and attributed to the
Boer-Mulders function.
The first measurement of the $\cos 2 \phi$
dependence of the proton-induced Drell-Yan process was recently reported for
$p+p$ and $p+d$ interactions
at 800 GeV/c~\cite{zhu07}. In contrast to pion-induced Drell-Yan,
significantly smaller (but non-zero) cos$2\phi$ azimuthal angular dependence
was observed in the $p+p$ and $p+d$ reactions. While the 
pion-induced Drell-Yan process
is dominated by annihilation between a valence antiquark in the pion
and a valence quark in the nucleon, the
proton-induced Drell-Yan process involves a valence quark in the proton
annihilating with a sea antiquark in the nucleon. Therefore, the
$p+p$ and $p+d$ results suggest~\cite{zhu07,zhang08} that 
the Boer-Mulders functions for
sea antiquarks are significantly smaller than those for valence quarks.

\section{Future Prospects at Fermilab and J-PARC}

Future fixed-target dimuon experiments have been proposed at
the 120 GeV Fermilab Main Injector and the 50 GeV J-PARC
facilities. As discussed earlier, the Fermilab E906 experiment
will extend the $\bar d/ \bar u$ asymmetry measurement to
larger $x$ region. Another goal of this experiment is to determine
the antiquark distributions in nuclei at large $x$ using nuclear
targets. New information on the quark energy loss in nuclei is 
also expected. An 
advantage of lower beam energies is that a much more sensitive study of the
partonic energy loss in nuclei could be carried out using the Drell-Yan
nuclear dependence~\cite{garvey03}.

With the possibility to accelerate polarized proton beams 
at J-PARC~\cite{p24},
the spin structure of the proton can also be investigated with the
proposed dimuon experiments. In particular, polarized Drell-Yan process
with polarized beam and/or polarized target at J-PARC would allow a unique
program on spin physics complementary to polarized DIS experiments
and the RHIC-Spin programs. Specific physics topics include the measurements
of T-odd Boer-Mulders distribution function in unpolarized Drell-Yan,
the extraction of T-odd Sivers distribution functions in singly
transversely polarized Drell-Yan, the helicity distribution of
antiqaurks in doubly longitudinally polarized Drell-Yan, and the
transversity distribution in doubly transversely polarized
Drell-Yan. It is worth
noting that polarized Drell-Yan is one of the major physics program at the
GSI Polarized Antiproton Experiment (PAX). The RHIC-Spin program will
likely provide the first results on polarized Drell-Yan. However, the
high luminosity and the broad kinematic coverage for the
large-$x$ region at J-PARC would allow some unique measurements to be
performed in the J-PARC dimuon experiments.

\bibliographystyle{aipproc}

\end{document}